\documentclass[11pt,preprint]{aastex}
\usepackage{epsfig,lscape}
\slugcomment{KSUPT-01/6 \hspace{0.5truecm} October 2001}

\def\deg{^{\circ}}

\def\dg{\nobreak^\circ}
\begin{document}

\title{Galactic Foregrounds in OVRO and UCSB South Pole 1994 Cosmic Microwave Background Anisotropy Data}

\author{Pia Mukherjee\altaffilmark{1}, Brian Dennison\altaffilmark{2}, Bharat Ratra\altaffilmark{1}, John H. Simonetti\altaffilmark{2}, Ken Ganga\altaffilmark{3}, \\ and Jean-Christophe Hamilton\altaffilmark{4,5}}

\altaffiltext{1}{Department of Physics, Kansas State University, 116 Cardwell Hall, Manhattan, KS 66506; pia, ratra@phys.ksu.edu.}
\altaffiltext{2}{Martin Observatory, Institute for Particle Physics and Astrophysics, and Department of Physics, Virginia Polytechnic Institute and State University, Blacksburg, VA 24061; dennison@astro.phys.vt.edu, jhs@vt.edu.}
\altaffiltext{3}{Infrared Processing and Analysis Center, California Institute of Technology, Pasadena, CA 91125; kmg@ipac.caltech.edu.}
\altaffiltext{4}{Physique Corpusculaire et Cosmologie, Coll\`ege de France, 11 Place Marcellin Berthelot, 75231 Paris, France.}
\altaffiltext{5}{Institut des Sciences Nucl\'eaires, 53 Av. des Martyrs, 38026 Grenoble, France; hamilton@in2p3.fr.}

\begin{abstract}
We study Galactic emission foreground contamination of the OVRO and UCSB South Pole 1994 cosmic microwave background anisotropy data by cross-correlating with templates of infrared dust emission and new high resolution VTSS and SHASSA $H_{\alpha}$ data.  $H_{\alpha}$ data provide rough upper limits on the level of free-free emission in the data sets.  The cross-correlation analysis does not contradict a two component foreground emission hypothesis, with the two dust-correlated components being free-free emission and spinning dust emission.
\end{abstract}

%\keywords{cosmology: observation --- cosmic microwave background ---% diffuse radiation --- ISM: dust, extinction}
   
\section{Introduction}
Cosmic microwave background (CMB) observations have begun to provide interesting constraints on cosmological parameters (see, e.g., Ratra et al. 1999a; Rocha et al. 1999; Knox \& Page 2000; Wang \& Mathews 2002; Podariu et al. 2001; Netterfield et al. 2002; Pryke et al. 2002; Stompor et al. 2001; Wang, Tegmark, \& Zaldariaga 2002; Efstathiou et al. 2002).  To determine high precision constrains from CMB anisotropy data, it must be carefully cleaned of any contaminating foregrounds to extract purely cosmological fluctuations to high accuracy.

So far the main sources of diffuse Galactic microwave emission known to be present are thermal dust, synchrotron, and free-free.  The identification and removal of free-free emission is not straightforward as it is likely to dominate the other two sources only in a small range of frequencies (around 60 GHz).  Galactic $H_{\alpha}$ line emission, which is produced by the same warm ionized medium (WIM) responsible for free-free emission, can be a good tracer of free-free emission at high Galactic latitudes where there is expected to be minimal extinction from dust.  Recently it has been proposed that there is at least indirect evidence for the presence of another component of Galactic emission, spinning dust emission from rotationally excited small dust grains.  The model of Draine \& Lazarian (1998a,b) predicts a microwave emission spectrum for this component that peaks at low frequencies ($\sim$ 10$-$30 GHz), with the details of the spectrum, such as the location of the peak, dependent on properties of the dust grains and their environment.  Specialized observations directed towards picking out this component are being conducted and the first tentative detection has been reported in Finkbeiner et al. (2002).

Indirect evidence for this component comes from microwave-infrared (hereafter IR) correlations.  Estimates of the correlation slope $a$ (defined in $\S$ 3) between 100 $\mu$m dust emission and microwave emission obtained from DMR (Kogut et al. 1996a,b), 19 GHz (de Oliveira-Costa et al. 1998), Saskatoon (de Oliveira-Costa et al. 1997), OVRO (Leitch et al. 1997), UCSB South Pole 1994 (Hamilton \& Ganga 2001) and Tenerife (de Oliveira-Costa et al. 1999; Mukherjee et al. 2001, de Oliveira-Costa et al. 2002) observations are such that they are roughly consistent with a free-free emission spectral index over this wide range of frequencies (10$-$90 GHz) and angular scales ($7\dg$ to $7^{\prime}$).  MAX 4 data show some IR correlation at 100 GHz in one field (Lim et al. 1996; Ganga et al. 1998).  No IR correlations were found in the Python V (Coble et al. 1999) or QMAP (de Oliveira-Costa et al. 2000) data.  (If the component that correlates with the IR dust continuum has similar scale dependence as IR dust emission itself, which has a multipole space power spectrum $\propto \ell^{-3}$, Kogut et al. 1996b, then this relation will be independent of scale\footnote{Bartlett \& Amram (1998) point out that such a power spectrum implies that the true sky variance remains roughly constant towards small scales, but estimates of this variance in restricted patches of sky may be ``pulled'' far from the true variance by important correlations on scales of the patch size.}.)   However, if this correlated emission were really free-free emission, then a similar correlation should exist between $H_{\alpha}$ and IR emission\footnote{The relation between free-free and $H_{\alpha}$ emission is $T_{\rm{ff}} \sim 7\ \mu {\rm{K}} \left(\frac{T}{10^4\ {\rm{K}}}\right)^{0.55, 0.85}\left(\frac{\lambda}{1\ {\rm{cm}}}\right)^{2.1}\left(\frac{I_{H_{\alpha}}}{{\rm{R}}}\right)$, at wavelength $\lambda$, where $T_{\rm{ff}}$ is the brightness temperature for free-free emission, $I_{H_{\alpha}}$ is the intensity of $H_{\alpha}$ emission, and the power law index on the gas temperature changes from 0.55 for temperature $T\leq 2.6\times 10^4$ K to 0.85 for $T>2.6\times 10^4$ K, so that at high temperatures there is a greater contrast in the levels of free-free and $H_{\alpha}$ emission.  The assumptions that go into deriving this relation include a fraction of singly ionized helium of unity, a helium to hydrogen ratio of 0.08, an approximate expression for the Gaunt factor, and further the relation is valid only for lines of sight in which interstellar extinction is negligible (Reynolds \& Haffner 2002).}.  So far $H_{\alpha}$-IR correlations have been found to be marginal.  Correlation slopes obtained (Reynolds et al. 1995; Kogut 1997; McCullough 1997; de-Oliveira-Costa et al. 2002) are a factor of a few lower than what is expected based on the microwave-IR correlation slopes and have significant uncertainties at present.  However, if the microwave-IR correlations are only in part due to free-free emission, then this may be acceptable.  This ratio may also vary across the sky due to the variation in the Hydrogen ionization fraction an
d optical extinction along different lines of sight (de Oliveira-Costa et al. 2002).  See Smoot (1998), Bartlett \& Amram (1998), McCullough et al. (1999), Valls-Gabaud (1998), Kogut (1999), and Draine \& Lazarian (1999) for reviews of the relation between $H_{\alpha}$ and free-free emission and for discussions of how spinning dust emission might be responsible for the ``anomalous'' IR-correlated emission.

While the cause of microwave-IR correlations is still under discussion, we present a correlation analysis of the OVRO and SP94 microwave data with templates of dust emission including the 12 $\mu$m template, and new $H_{\alpha}$ templates, to see whether we are able to separate free-free emission and detect spinning dust emission.  These experiments are sensitive to very different angular scales and span a wide frequency range ($\sim$ 15 GHz to 40 GHz) where a considerable contribution from spinning dust emission may be expected.  The foreground templates are introduced in $\S$ 2, the method is outlined in $\S$ 3, the OVRO foregrounds analysis is presented in $\S$ 4, the SP94 foregrounds analysis is presented in $\S$ 5, and $\S$ 6 consists of an overall discussion and conclusions.

\section{The Foreground Templates}

We use the 100 $\mu$m template (Schlegel, Finkbeiner, \& Davis 1998) as a tracer of emission from large dust grains which are in equilibrium with the interstellar medium and radiate thermally at temperatures around 20 K.  The 60 and 12 $\mu$m templates (D. Finkbeiner, private communication 2000) are used as tracers of emission from smaller dust grains that are transiently heated to higher temperatures and emit (thermally at these high frequencies and rotationally at microwave frequencies) to get back to their ground state.  That the 12 $\mu$m template is a clean tracer of emission from small dust grains and that such emission has a detectable counterpart at microwave frequencies is still being tested.  Even if the 12 $\mu$m template is a clean tracer of the distribution of small dust grains, it may not be a good tracer of spinning dust emission at microwave frequencies which depends not only on the distribution of grains but on various other local factors.  We proceed with these caveats in mind.  It is thought that the distributions of small and large dust grains may be similar on large scales, but in some regions of the sky their small scale distribution could possibly differ (Weingartner and Draine 2000).  Thus, as noted by de Oliveira-Costa et al. (2002), shorter wavelength dust template maps might be expected to trace spinning dust at least as well as the 100 $\mu$m map and possibly better on small scales.  The templates are labelled i100, i60 and i12 in the tables that follow.   These templates are such that between 100 $\mu$m and 60 $\mu$m emission decreases over $90\%$ of the sky and between 60 $\mu$m and 12 $\mu$m the emission increases over $60\%$ of the sky. Over all, between 100 $\mu$m and 12 $\mu$m the emission increases over about $10\%$ of the sky at high Galactic latitudes.  Emission from large dust grains is expected to drop to insignificant levels at 12 $\mu$m.  In spatial structure the 60 $\mu$m template is very similar to the 100 $\mu$m template over most parts of the sky.

When cross-correlating the OVRO data with dust templates, cross-correlating with the dust column density template (a template of the amount of thermal dust emission expected if the dust were all at a reference temperature of 18 K, Schlegel et al. 1998, and is labelled cdi100 in the tables) is necessary in order to make the estimates of the correlation slope $a$ (defined in $\S$ 3) comparable to those from the DMR data say, as the temperature in the NCP region, and especially in the OVRO fields, is lower than the global average, resulting in lower dust emission and a correspondingly higher inferred correlation coefficient (the factor is 2.7 for OVRO, this was emphasized by Finkbeiner \& Schlegel 1999).  Table 1 lists the dust temperatures and intensities in the OVRO fields, the whole of the NCP region (DEC $>81\dg$), and the high Galactic latitude sky ($|b|>30\dg$).  The estimate from the Saskatoon analysis may need to be revised similarly.  Note also that there is greater dust emission in the OVRO fields and this can play an important role in heating diffuse ionized gas.

The $H_{\alpha}$ observations for the OVRO region were carried out on October 4, 2000 with the Virginia Tech Spectral Line Imaging Camera (SLIC) (Simonetti, Dennison, \& Topasna 1996; Dennison, Topasna, \& Simonetti 1997). The SLIC uses
a 1.7 nm bandpass filter centered on $H_{\alpha}$ and a dual bandpass
continuum filter with narrow passbands on either side of the $H_{\alpha}$
line. Both filters reject the [NII] lines at 654.8 and 658.3 nm. An
$f/1.2$ Noct-Nikkor lens images a 5-degree radius circular field on a
cyrogenically cooled TK $512 \times 512$ CCD camera. A separate
auto-tracking CCD camera provides arcsecond precision tracking. The
facility is located at a dark site at the Virginia Tech Horton Research
Center.

We obtained seven 6-minute $H_{\alpha}$ images and six 3-minute continuum
images approximately centered on the North Celestial Pole (NCP). We also
recorded ten continuum flat field images using a flat field light box
described in Simonetti et al. (1996). The images were combined and flat corrected in IRAF\footnote{IRAF is distributed by the National Optical Astronomy
Observatory, which is operated by the Association of Universities for
Research in Astronomy, Inc. under cooperative agreement with the
National Science Foundation.} using standard techniques (Dennison et al. 1997).
For off-axis rays the bandpass of the interference filter shifts to
longer wavelengths limiting the usable field to a 5-degree radius. The
sensitivity to $H_{\alpha}$ is quite flat over this field, but was radially
corrected using $H_{\alpha}$ observations of a flat $H_{\alpha}$ light box.  
$H_{\alpha}$ calibration was achieved through repeated
 observations of the bright planetary nebula M76, which we have
 previously determined to have an $H_{\alpha}$ flux of $7.95 \times
 10^{-11}$ ergs s$^{-1}$ cm$^{-2}$.  The $H_{\alpha}$ image was then continuum-corrected through a weighted subtraction
 of the continuum image. This first required offsetting the continuum
 image by 1.93 pixels to achieve registration. (The different filters
 exhibit slightly different transverse gradients of optical thickness and
 thus displace the images by different amounts.) The $H_{\alpha}$ image was
 then smoothed by 0.5 pixel to match the resolution of the offset
 continuum image. The continuum-corrected image has significantly reduced
 stellar contamination, although the brightest stars are not well
 subtracted due to saturation effects.

The $H_{\alpha}$ line lies within a band of atmospheric OH emission, and because of the bandpass shift with incidence angle, our image contains contamination from the OH(6-1)$P_2(4)$ line at 656.9nm in the center of the image and the OH(6-1)$P_1(3)$ line at 655.4 nm \citep{Ch} in the outer part. The latter emission is manifested as a faint ($\approx 3$ R), diffuse ring with peak radius $\approx 4.\!^\circ 6$. An empirical model ring derived from azimuthally smoothed observations of a high-Galactic-latitude field devoid of bright stars was used to remove the ring in the NCP field. The fit was optimized to remove the ring between $1.\!^\circ6$ and $3.\!^\circ0$, the region of interest in this study.
 Some contamination remains outside of this range. The bright star, Polaris, may also contribute to the contamination in the center of the image. A uniform gradient caused by scattered light as well as Galactic starlight was also removed from the image.  
Because the image contains an unknown, nearly uniform background due to geocoronal $H_{\alpha}$ emission and scattered light, the median value of the final image was set equal to zero.

Coordinates were established using WCSTools (Mink 1997) which was used to fit to stars from the Hubble Guide Star Catalog.  The final image is calibrated in rayleighs (R), where 1 R = $10^6/4\pi$ photons cm$^{-2}$ s$^{-1}$ sr$^{-1}$.

For the SP94 region the recently released Southern $H_{\alpha}$ Sky Survey (SHASSA) $H_{\alpha}$ observations (Gaustad et al. 2001) were used.

\section{The Method}
The method of cross-correlating microwave data with available foreground emission templates was pioneered by Kogut et al. (1996a,b), G\'orski et al. (1996) and Banday et al. (1996) and further generalized by de Oliveira-Costa et al. (1997). It assumes that the microwave data consist of a super-position of 
CMB anisotropy, noise, and foreground emission components,
\begin{equation}
y = aX + x_{CMB} + n.
\end{equation} 
Here $y$ is a data vector of $N$ pixels, $X$ is an $N\times M$ 
element matrix containing $M$ foreground emission templates convolved with the 
experimental beam and processed by the observing strategy adopted, and $a$ is a vector of size $M$ that represents 
the levels at which these foreground templates are present in the data 
(each component of $a$ is thus the correlation slope for the corresponding template).  The vector $n$ is the noise in the data and $x_{CMB}$ is the CMB anisotropy convolved with the beam and processed by the observing strategy of the experiment. The noise and CMB are treated as mutually uncorrelated.
 
The minimum variance estimate of $a$, obtained by minimizing $\chi^2=(y-aX)^TC^{-1}(y-aX)$, is 
\begin{equation}
\hat{a} = [X^{T} C^{-1} X]^{-1} X^{T} C^{-1} y,
\end{equation} 
with errors $\Delta \hat{a}_i = \Sigma_{ii}^{1/2}$ 
where the matrix $\Sigma$ is
\begin{equation}
\Sigma = <\hat{a}^2> - <\hat{a}>^2 = [X^{T} C^{-1} X]^{-1}. 
\end{equation}
Here $C$ is the total covariance matrix (the sum of the noise covariance matrix and the CMB covariance matrix).  In these equations $X$ and $y$ are actually deviations of the corresponding quantities from the weighted mean with weights given by $C^{-1}$.  The rms amplitude of temperature fluctuations in the data that results from the correlation is $\Delta T = (\hat{a}\pm \delta\hat{a}) \sigma_{\rm{fore}}$, where $\sigma_{\rm{fore}}$ is the rms deviation of the corresponding foreground template map.

Underlying this method of cross-correlating CMB data with maps of Galactic emission are the assumptions that the templates perfectly trace the spatial appearance of the respective components at the frequencies and angular scales of the CMB data, and that the templates are exhaustive, i.e., they are sufficient to explain all the structure in the data.  In addition, for deriving the minimum variance estimator, we assume that the noise has zero mean and a Gaussian distribution.\footnote{As noted in Mukherjee et al. (2001), violation of the assumptions that the templates are perfect and exhaustive can be expressed as an additional ``missing'' template which would by construction be perfectly compensating. 
If this template is uncorrelated with the templates used and has zero mean, we would expect only an increase in the noise level.  If the template is correlated or has non-zero mean the estimate of  the correlation slope $a$ would be biased.}  With such caveats in mind, this method has been widely used to test for the presence of contaminants in CMB anisotropy data.

\section{The OVRO Correlation Analyses}
\label{micdust}

The OVRO observations consist of 36 fields around the NCP, at a declination of $\sim 88\dg$, at frequencies of 14.5 and 31.7 GHz, with a double differencing $7'$ FWHM beam (see Leitch et al. 2000 for more information on the OVRO experiment).  When the observed data are modelled as consisting of a CMB component and a single foreground of spectral index $\beta$, $\Delta T_i(\nu)=\Delta T_{{\rm{cmb}},i} + \Delta T_{{\rm{fore}},i} \nu^{\beta}$, then given observations at two frequencies, one can solve for the CMB anisotropy component in terms of the unknown spectral index of the foreground,
\begin{equation}
\Delta T_{{\rm{cmb}},i}=\frac{\Delta T_i(\nu_1)\nu_1^{-\beta}-\Delta T_i(\nu_2)\nu_2^{-\beta}}{\alpha(\nu_1) \nu_1^{-\beta} - \alpha(\nu_2) \nu_2^{-\beta}} ,
\end{equation}
where $\alpha(\nu)=x^2 e^x/(e^x-1)^2$, with $x=h\nu/kT$, is the correction factor for the RJ approximation to a black body spectrum.  We use this inferred CMB signal and perform a likelihood analysis.  Here the total covariance matrix is composed of the CMB anisotropy flat bandpower theory covariance matrix, the instrumental noise matrix, and the (one-off diagonal) correlated noise $\sigma_n=41$ $\mu$K found in the 31.7 GHz data (Leitch et. al. 2000).  This process assumes Gaussianity of the inferred CMB anisotropy signal and the presence of only a single foreground.  A constraint matrix to account for the removal of a constant offset is included in the analysis.  After marginalizing over $\beta$ and $\sigma_n$, the constraints on the level of CMB anisotropy agree with those found by Leitch et al. (2000).  The constraints on $\beta$ are not stringent with $\beta>-1.8$ ruled out at 2 $\sigma$. Although the data are unable to rule out a very steep spectrum foreground, based on correlating the data with Westerbork Northern Sky Survey (WENSS, Rengelink et al. 1997) 325 MHz maps, Leitch et. al. (1997) conclude that the contribution of any $\beta<-2.2$ spectrum component is negligible.  This then leaves a significant amount of foreground at 14.5 GHz of $\Delta T_{\rm{rms}} \sim 175$ $\mu$K, though it indicates that the 31.7 GHz data are almost entirely CMB anisotropy.

Table 2, in its upper panel, lists results from correlating the OVRO data with each dust template individually.  Of the three templates of dust emission, the 100 $\mu$m template correlates best with the 14.5 GHz data, followed closely by the 60 $\mu$m template, and then the 12 $\mu$m template, all at a high level of significance.  The reverse trend may be true for the 31.7 data, however the significance of dust correlation with the 31.7 GHz data is small\footnote{If we calculate the correlation slope after marginalizing over $\sigma_n$, the correlation slope so found is well within the error bars of the numerical values in Table 2 but with slightly larger error bars.}.  The correlation slope for the cdi100 template is a factor of about 3 lower than that for the i100 template.  The results are independent of the input CMB power spectrum.  The $\chi^2$ of the fits indicate that the small non-CMB signal in the 31.7 GHz data, inferred from eq. (1), is satisfactorily accounted for, and about half of the non-CMB signal at 14.5 GHz can be accounted for by the dust templates\footnote{We have tested for possible biases in this minimum variance analysis method by analyzing 1000 simulations of the data.  These were made by simulating the CMB anisotropy component from a model and adding noise and a certain level of foreground on to it.  We find that the weighted mean of the res
ulting correlation slopes turns out to be what was put in to a high level of accuracy and the weighted rms deviation of the correlation slopes from the mean was in good agreement with the error bar for the chosen model.  Thus not only is the method not biased, but the error bars are also accurate.}.  However, the inferred foreground and the fitted foreground have similar spatial structure so it is possible that dust-correlated emission might satisfactorily explain all the foreground in the 14.5 GHz OVRO data and the discrepancy might be due to uncertainties in the foreground template.  If the templates are assumed to be close to perfect, then a better job can be done if we are able to cleanly separate the two components of emission likely to be present in the data, as discussed below.  Another possibility is that there is an unaccounted-for foreground, possibly free-free emission uncorrelated with dust.

The central panel of Table 2 lists results from correlating the OVRO data sets with two foreground templates jointly.  When correlating jointly with the 12 $\mu$m and 100 $\mu$m templates, all the correlation in the 14.5 GHz data gets attributed to the 100 $\mu$m template.  At 31.7 GHz there might be slight evidence of the emission getting attributed to the higher frequency template.   These dust templates have similar spatial structure in the OVRO fields ($r=\Sigma^{-1}_{ij}/(\Sigma^{-1}_{ii} \Sigma^{-1}_{jj})^{0.5}$, the weighted correlation coefficient between the templates indexed $i$ and $j$, is 0.7, see Figure 1).  When the 60 $\mu$m and 100 $\mu$m dust templates are correlated jointly with the OVRO data, the high correlation between these templates ($r=0.9$) results in large uncertainty in separating the two dust emission components.  (When the 12 $\mu$m and 60 $\mu$m templates are jointly correlated the result is similar to when the 12 $\mu$m and 100 $\mu$m templates are jointly correlated.)

If the dust-correlated emission were entirely spinning dust emission, then a stronger correlation with the higher frequency dust templates should be expected.  The results from individual and joint template correlations demonstrate that this is not the case for the OVRO 14.5 GHz data.  The 12 $\mu$m-correlated emission is subdominant to the 100 $\mu$m-correlated emission in the 14.5 GHz OVRO data, even though the two templates have similar spatial structure.  This may happen if the 12 $\mu$m template is a better tracer of spinning dust emission, and the 100 $\mu$m template additionally traces free-free emission.  The numbers in Table 2 would then indicate that there are comparable amounts of the two components present in the OVRO data at 14.5 GHz.

More microwave data should be used to check this deduction, which is only tentative at present given the caveats discussed in the earlier section and the shortcomings and possibilities for systematic errors in this correlation method discussed in detail in Mukherjee et al. (2001) and mentioned in section $\S$ 3.  If however this deduction is true then the 14.5 GHz data are consistent with about $92\pm 16$ $\mu$K rms of spinning dust emission and hence, using the 100 $\mu$m-correlated emission results for the 14.5 GHz data (Table 2), with roughly $101\pm22$ $\mu$K rms of free-free emission at 14.5 GHz.  This would translate into $19\pm4$ $\mu$K rms of free-free emission at 31.7 GHz.  Note, however, that the uncertainties in estimates of the correlation slope and the corresponding temperature fluctuations are comparable at the two frequencies.  Hence, since the dust-correlated foreground components are present only at low levels at 31.7 GHz the significances of the correlations are low as expected, becoming even lower when the data at this frequency are correlated with the 12 $\mu$m and 100 $\mu$m templates together.  This makes it hard to ascertain which component dominates or even distinguish between two components at 31.7 GHz.   However the estimated $a$ values do not contradict the above scenario, which would explain why the spectral index of the 100 $\mu$m-correlated emission, $-2.6^{+0.9}_{-2.0}$, is more skewed towards steeper values while being consistent with free-free, since the 100 $\mu$m-correlated emission at 14.5 GHz contains in it a 12 $\mu$m correlated component as well, which itself has a spectral index of $-1.7^{+0.9}_{-1.5}$ between the two frequencies, probably consistent with what is expected for spinning dust emission.

When the $H_{\alpha}$ data are convolved with the OVRO beam and double differenced, the data have an rms of 0.83 R in the OVRO fields.  To find the visual extinction (which results from a combination of scattering and absorption) expected in the OVRO fields, we use the reddening maps provided by Schlegel et al. (1998) and find that the average reddening in the 36 OVRO fields is E(B-V) = 0.297 mag, implying a high visual extinction of 0.921 mag (using R(V) = 3.1).\footnote{ The average extinction in the whole of the NCP (DEC $>81\dg$) region is 0.15 mag, in the $b>30\dg$ region it is 0.04 mag, and in the $b<-30\dg$ region it is 0.05 mag.}  This implies that the $H_{\alpha}$ signal seen is at most $10^{-0.4\times 0.921}=0.428$ or $43\%$ of the total $H_{\alpha}$ signal, or that the total $H_{\alpha}$ signal is 2.3 times that observed.  This correction for extinction is not accurate as the decrease in the $H_{\alpha}$ surface brightness in a given direction depends not only on the total column density of dust but also on the unknown arrangement of the emitting and absorbing material along the line of sight (Reynolds \& Haffner 2002).  Without attempting to make a more accurate correction for extinction, 0.83 R would imply an estimated $H_{\alpha}$ signal of 1.9 R and a free-free signal of $63$ $\mu$K rms at 14.5 GHz.  This level of free-free emission estimated from $H_{\alpha}$ data is an upper limit because some of the $H_{\alpha}$ signal could be due to incompletely subtracted stars.\footnote{If, in an attempt to remove stellar residuals, we set, prior to convolution and double differencing, the signal in pixels in the $H_{\alpha}$ map that show structure greater than $3$ $\sigma$ to $\sigma$ which is the local rms deviation from the local mean, where local here is taken to be from half a degree below to half a degree above the OVRO ring, then the rms of the resulting $H_{\alpha}$ signal in the 36 fields drops to 0.39 R (0.48 R before the double differencing).  We have repeated the analysis with this template and the results and conclusions presented here still hold.}

Thus the upper limit on free-free emission inferred from $H_{\alpha}$ data is less than the free-free like 100 $\mu$m-correlated signal found above, and is much less than the signal in the OVRO data at 14.5 GHz that was found to have a spectral index like that of free-free emission pointed out at the beginning of this section.  Could there be a preferential heating of the WIM, given that gas gets heated more easily than dust, due to the excess dust emission in this region, such that the $H_{\alpha}$ data could be used to deduce a higher level of free-free emission in this region?  For there to be, say 100 $\mu$K rms of free-free emission at 14.5 GHz, there needs to be about 3 R of (correlated) $H_{\alpha}$ fluctuations, or only 1.3 R in the presence of $43\%$ visual extinction.  If the gas temperature here were higher by a factor of 2.3, then the level of $H_{\alpha}$ fluctuations expected can come down to 0.83 R.\footnote{Leitch et al. (1997) conclude that the temperature needs to be higher by a factor $\sim 100$ in order to explain the level of foreground in the OVRO data.  This is because they wanted to explain a free-free emission signal of 300 $\mu$K (a factor of 3 larger from what is argued for here), they had not taken the high visual extinction in the OVRO fields into account (this causes a difference of a factor of 2.3), and had an $H_{\alpha}$ signal of rms $<0.1$ $R$ (factor of 4 lower at least from what is found here), giving in total a difference of a factor of 28 from that found here, which when multiplied by 4.2 leads to the factor $~\sim 100$.}

The results from correlating the OVRO data with the $H_{\alpha}$ template are given in the bottom panel of Table 2.  When correlating the $H_{\alpha}$ template with microwave data one may not find an observable correlation when there is significant visual extinction due to dust grains for which the $H_{\alpha}$ template has not been corrected, and also when the data contains a significant amount of another foreground.  However, a negative correlation with the $H_{\alpha}$ template is found when the 14.5 GHz data are correlated jointly with the $H_{\alpha}$ and 100 $\mu$m or 12 $\mu$m templates with somewhat greater significance being attributed to the dust correlations than was obtained in the case of individual correlations.  From this we deduce that the structure in the $H_{\alpha}$ template partially traces the structure in the dust templates, sky-rotations indicate the same.  This can also be seen from Figure 1.  Note however that this correlation becomes apparant only after the OVRO double differencing.  Before the OVRO double-differencing, none of the dust templates show a significant correlation with the $H_{\alpha}$ template.  Low $H_{\alpha}$-dust correlations may be acceptable, however, due to various reasons\footnote{Ionized gas emits $H_{\alpha}$ and free-free in direct proportion, with the surface brightness of emission proportional to the electron density squared.  These emissions only indirectly trace thermal emission from the dust associated with the ionized medium which has a surface brightness proportional to the density of dust.  Further dust absorbs $H_{\alpha}$ light and so about $25\%$ of high latitude $H_{\alpha}$ emission is estimated to be scattered light (McCullough 1997).  Lagache et al. (2000) have used $H_{\alpha}$ and HI data to decompose the far infrared thermal dust emission at high Galactic latitude into components associated with the WIM and the warm neutral medium (WNM).  They confirm that about $20-30 \%$ of thermal dust emission is associated with the WIM.  The correlation between ionized gas and dust emission may also be angular scale dependent (McCullough 1997).}, including the possibility that microwave-dust correlations are only in part due to free-free emission, in patches of the sky with high gas temperature and optical extinction, and on small angular scales.  Since all these are true of the OVRO fields, estimates of correlations involving $H_{\alpha}$ data in the OVRO fields may be suspect and we do not present numbers for $H_{\alpha}$-dust correlations here.

\section{Foreground Emission in the UCSB South Pole 1994 Data}

The SP94 data were taken in two bands.  The $K_a$ band had four channels (centred at 27.25, 29.75, 32.25, and 34.75 GHz), hereafter $K_a$1, $K_a$2, $K_a$3, and $K_a$4, and the $Q$ band had three (at 39.15, 41.45, and 43.75 GHz), hereafter $Q$1, $Q$2, and $Q$3.  The beams had resolution of about $1\dg$ and there are 43 data points in each frequency channel, all at a declination of $-62\dg$, with RA ranging from $23\dg$ to $67\dg$.  The data were obtained using a sinusoidal $1.\!^\circ5$ chop with smooth, constant declination, constant velocity scans.  See Gundersen et al. (1995) and Ganga et al. (1997) for more details on the SP94 experiment.  Hamilton \& Ganga (2001) have reported the detection of 100 $\mu$m-correlated emission at 1.6 $\sigma$ confidence in the $Q$ band data, resulting in a possible reduction of the deduced CMB anisotropy band temperature of Ganga et al. (1997).  The $K_a$ band data was found to be free of contamination from this template.

The SP94 cross-correlation analysis here is done ignoring the last 4 SP94 pixels that are close to and affected by a set of pixels at RA $\sim 69^\circ$ that have negative values of flux in the 12 $\mu$m map, apparently due to inaccurate source subtraction at that location.\footnote{If we include all the SP94 pixels in the analysis, then we get a strong correlation of the SP94 data with the 12 $\mu$m template.  The pixels in the 12 $\mu$m map that have negative flux values cause a dip in the signal in the last few pixels of the SP94 beam convolved template (even when the negative flux values are set to 0 or to a value lower than the local mean, assuming that we can not really know the level of diffuse emission here due to the presence of a source that has not been accurately removed).  There is a similar dip in the microwave signal in the different channels of the SP94 data (see Figure 2), and this results in a strong correlation with the 12 $\mu$m template.}  We show the results of correlating the data with the 100 $\mu$m and the 12 $\mu$m templates individually because when the data are correlated with these templates jointly, then due to the high correlation between the templates ($r=0.8$, see Figure 2), uncertainties increase further and nothing can be gained, given the already low level of contamination in this data.  Estimates of the correlation slope are derived by jointly correlating all channels of the $K_{a}$ band, the $Q$ band and also all channels of both bands together, following Hamilton \& Ganga (2001).  This is done because the anisotropy model covariance matrix causes non-trivial covariance between pixels of all channels.  The noise in a given pixel is correlated with noise in the same pixel in other channels of the same band.  The noise in the $K_a$ and $Q$ bands are uncorrelated.  We then combine the $a$ values obtained for each template into a representative (average) number ($\bar{a}$) using both a spectral index $n$ (here $\bar{a}=\bar{a_0}(\nu/\nu_0)^n$, where $\nu_0$ is a reference
 frequency) of 0 and the best fit $n$, as described in Hamilton \& Ganga (2001), for the $K_a$ and $Q$ bands separately and for both of them together.  The best estimate of $\bar{a}$ is obtained, in the case $n=0$, from
\begin{equation}
\bar{a}=\frac{\rm{Total}(\Sigma^{-1}a)}{\rm{Total}(\Sigma^{-1})}
\end{equation} 
where Total denotes the sum of all elements of a matrix or vector.  This is thus a weighted average taking account of correlations between the $a$ values of the different channels.  The best fit $n$ is obtained by minimizing, over a grid of values of $n$, $\chi^2=(a-\bar{a_0}m)^T\Sigma^{-1}(a-\bar{a_0}m)$, where $m$ is a vector with elements $m_i=(\nu_i/\nu_0)^n$.

Results are shown in the Table 3.  We see that the $K_a$ band data does not correlate with the 100 $\mu$m template while the $Q$ band data correlates marginally, the best fit spectral index when both bands are considered together is positive, and there is a greater than 2 $\sigma$ detection of 100 $\mu$m-correlated emission, in agreement with Hamilton \& Ganga (2001).  Regarding 12 $\mu$m-correlated emission, we again find that the $K_a$ band data does not correlate while the $Q$ band data correlates marginally, the best fit spectral index when both bands are considered together is positive, and there is a greater than 2 $\sigma$ detection of 12 $\mu$m-correlated emission when all channels are considered together with the best fit spectral index.  The upper limit to the total contamination from dust templates in the SP94 data is $\Delta T_{\rm{rms}}=26\pm 11$ $\mu$K when the results of all channels are combined using estimates of the best fit spectral indices.  This is an upper limit because the 100 and the 12 $\mu$m correlated emission have been coadded, even though these templates are seen to have similar structure (see Figure 2).  We are unable to say from this kind of a correlation analysis which of these two components is dominant or even whether these correlations are due to two physically distinct components. Most of the contamination in the SP94 data is in its $Q$ band (total dust-correlated emission in this band is at most $\Delta T_{\rm{rms}}=32\pm 21$ $\mu K$), as a consequence of which the deduced SP94 band temperature (Ganga et al. 1997) in the $Q$ band could go down in amplitude by at most $21\%$, to $41\pm26$ $\mu$K.  No clear evidence is found for the contamination of $K_a$ band data.  This is in agreement with the conclusions of Ganga et al. (1997), also see Ratra et al. (1999b).

Sky-rotations (correlating the data with template dust maps obtained by rotating the initial map around the Galactic poles and by inverting the North and South Poles) indicate that the correlations are as significant as the numbers indicate.  In other words the correlation that could arise from a random alignment between CMB anisotropies and the dust template has been taken into account in the analysis by the data and CMB covariance matrices, and the rms of the correlations from random patches agrees well with the error bars obtained by the correlation method.  Correlations similar to those obtained using the 100 $\mu$m template are obtained when the data are correlated with the column density template for 100 $\mu$m emission, indicating that the dust temperature in the SP94 pixels is not radically different from the average.  The optical extinction in this region is also small.

We analyze the structure of $H_{\alpha}$ signal in the SP94 pixels using data from SHASSA.  These data are available at a resolution of $0.\!^\circ1$.  No significant correlation is found when the SP94 data are correlated with the SP94 beam convolved $H_{\alpha}$ data.  When the SP94 data are correlated jointly with the 100 $\mu$m and $H_{\alpha}$ templates, a negative correlation is attributed to the $H_{\alpha}$ template ($K_{a}+Q$: $n=0$, $a=-3.6\pm4.1$ and $n=3$, $a=-3.9\pm2.1$) while a positive correlation is attributed to the 100 $\mu$m template ($K_{a}+Q$: $n=0$, $a=35.6\pm34.4$ and $n=2.2$, $a=72.7\pm22.8$), of greater significance than that obtained when the data were correlated with this dust template individually.  Thus, like in the OVRO case, we find that the structure in the $H_{\alpha}$ template has something in common with structure of dust emission (see Figure 2), and that in such a situation the correlation method is unable to separate any $H_{\alpha}$-correlated signal from the rest of the dust-correlated signal.  The correlation slope between the $H_{\alpha}$ signal and the 100 $\mu$m dust emission, for example for channel $Q$1, is $0.3\pm0.2$ R(MJy/sr)$^{-1}$ ($10\%$ of the rotated dust patches have greater correlation slope, $8\%$ of the dust patches correlated with greater significance).  The correlation coefficient is $0.4\pm0.2$ (the same $8\%$ of the dust patches had greater correlation coefficient than the real dust patch).  Here the variance is found by correlating the $H_{\alpha}$ data with 144 rotated versions of the dust emission template.  As in the case of OVRO, the $H_{\alpha}$ and dust data correlate better after the chopping strategy of the SP94 experiment has been simulated on these data. More sensible deductions regarding any such association can be made by comparing the structure in these maps over larger areas of sky (rather than in small 1d strips).

\section{Discussion and Conclusion}
Consider the scenario in which dust-correlated emission consists of free-free emission as well as another component.  The amplitude of dust-correlated emission in DMR data at 53 GHz is $\Delta T_{\rm{rms}}=7.1\pm 1.7$ $\mu$K, while the total amount of free-free emission is $\Delta T_{\rm{rms}}=5.2\pm 4.2$ $\mu$K using multifrequency data (Kogut et al. 1996b).  Since at that time no other dust-correlated component of emission was expected to be found at these frequencies, it was concluded that the correlated component forms at least a third (at $95\%$ confidence) of the total free-free emission and possibly even the bulk of it. If instead free-free emission were only a part of the total dust-correlated emission (with the 12 $\mu$m and 100 $\mu$m templates being similar in structure at large angular scales) then the remaining part is constrained to be $5\pm 4$ $\mu$K rms at 53 GHz on DMR scales.  If this component is independent of angular scale, then this estimate together with the amplitude of 12 $\mu$m-correlated emission at 14.5 GHz (OVRO) and 36 GHz (SP94) implies a spectral index of $\beta=-2.2^{+0.6}_{-1.4}$ between 14.5 GHz and 53 GHz and $\beta=-3.0^{+3.0}_{-5.1}$ between 36 GHz and 53 GHz (between 14.5 GHz and 31.7 GHz $\beta=-1.7^{+0.9}_{-1.5}$ as discussed in $\S$ 4).  These spectral index estimates may be consistent with the spinning dust hypothesis, becoming more negative at higher frequencies, although the uncertainties in the estimates are significant.  For free-free emission, if we use the power spectrum estimated by Veeraraghavan \& Davies (1997), $C_l \propto l^{-2.3}$ (they estimated the power spectrum directly from cleaned and destriped $H_{\alpha}$ maps and so may have obtained a more reliable estimate of the free-free emission power spectrum than did Kogut et al. 1996a), and predict the level of free-free emission that the OVRO 14.5 GHz and the SP94 36 GHz experiments should see at their respective angular scales given the average estimate of COBE DMR, we get $26\pm 20$ $\mu$K rms and $6\pm 5$ $\mu$K rms for OVRO and SP94 respectively.  This implies free-free emission of $< 66$ $\mu$K rms at $95\%$ confidence in the 14.5 GHz OVRO data, consistent with the rough upper limits indicated by $H_{\alpha}$ observations, though the emission in smaller regions could deviate significantly from this average.  We see that it is plausible that the 100 $\mu$m template, at least in some regions of the sky, traces free-free emission, as well as the structure in the 12 $\mu$m template by virtue of the large and small grains being well mixed.

If it is generally true that the 100 $\mu$m template correlations are in part free-free, the remaining being another component that the 12 $\mu$m template traces better (this may hold only in regions of the sky where large and small dust grains are well mixed, also our analysis throughout assumes that the emission from these grains are well traced by the templates we use which are thus close to perfect as discussed), this hypothesis would explain why Veeraraghavan \& Davies (1997) obtain a considerably lower normalization for the free-free emission power spectrum derived from $H_{\alpha}$ maps than did Kogut et al. (1996a), who used dust-correlated microwave emission (which is only in part free-free emission).   It would also explain the flatter spectrum of free-free emission that they obtained given that the remaining dust-correlated emission makes a contribution to temperature fluctuations that seems to be independent of angular scale, from the estimates available to date.

Figure 3$a$ is a plot of 100 $\mu$m-correlated emission found in microwave data so far.  It is similar to Fig 4. of Hamilton \& Ganga (2001) with the corrected OVRO points with error bars added.  Spectrally the data do not seem to favour either of the spinning dust or free-free interpretations.  Note that the set of Tenerife points taken from de Oliveira-Costa et al. (2002), for the region $|b|>20\deg$ no longer show a rising spectrum, the key signature of spinning dust emission, between 10 and 15 GHz (though they did find that microwave-$H_{\alpha}$ correlations were too small to explain the microwave-IR correlations in the data).  The points taken from Mukherjee et al. (2001) are higher because there it was shown that only pixels close to the Galactic plane actually show dust-correlated emission and here the estimated $a$ is higher.  For such reasons comparisons between experiments must be made with caution.    

Figure 3$b$ is a plot of 12 $\mu$m-correlated emission in microwave data.  The Tenerife points (taken from de Oliveira-Costa et al. 2002) could go up if the estimate is confined to the limited region that shows dust-correlated emission at both frequencies (Mukherjee et al. 2001).  The figure indicates that spinning dust emission might have been detected in microwave data with significant levels of emission at about 15 GHz.  We however need to check for example whether the correlation slope between the 12 $\mu$m template and DMR data lies on the spectral curve shown.  In order to correlate the DMR data with the 12 $\mu$m template we may need to await the cleaning of the template as the signal at 12 $\mu$m is contaminated significantly by zodiacal emission.  Modelling zodiacal emission is a complex task and this could take some time.  We were able use the 12 $\mu$m template in its present state as we consider only small regions of the sky, and in these few pixels far from the ecliptic plane there does not seem to be obvious contamination.  So, whether the 12 $\mu$m-correlated emission in other CMB anisotropy data sets follow the spectral pattern of this figure remains to be determined.  Higher quality data from the newer CMB experiments, such as CBI and DASI, and that anticipated from  MAP, should help resolve this issue.

Rather than having to use thermal dust emission as an indirect tracer of free-free emission, data from new $H_{\alpha}$ observations are expected to provide a template that more directly traces free-free emission.  However, converting the $H_{\alpha}$ data into a reliable template of free-free emission may require some care.  A careful comparison of structures in a high resolution $H_{\alpha}$ map, with information on extinction and gas temperatures (say from X-ray and ultraviolet observations), to structures in microwave and IR data over larger areas of sky is needed to better understand the relevant associations.  The spinning dust hypothesis itself is being studied by making pointed observations towards dust features predicted to emit significant amounts of spinning dust emission (Finkbeiner et al. 2002).  Such observations can more easily and significantly constrain the spinning dust model.  Even if these foreground contaminants are present only at low levels in CMB anisotropy data, it is important that they be understood and removed to high precision, especially if one wishes to study higher order statistics and non-Gaussian signatures in the CMB anisotropy data (see, e.g., Park et al. 1998, 2001; Chiang \& Coles 2000; Barreiro \& Hobson 2001; Wu et al. 2001; Park, Park, \& Ratra 2002, Shandarin et al. 2002).

In summary, we discuss a scenario in which 100 $\mu$m-correlated microwave emission in the OVRO and SP94 microwave anisotropy data sets consists of free-free emission as well as another component.  This is motivated by the facts that free-free emission by itself is not expected to result in strong microwave-IR correlations, free-free emission is constrained by $H_{\alpha}$ emission in both the data sets to be less that the amplitude of dust-correlated emission, and that the 100 $\mu$m-correlated emission dominates over the 12 $\mu$m-correlated emission in the OVRO 14.5 GHz data.  The $H_{\alpha}$ data and dust emission data seem to have some common structure, but dust emission correlates better with microwave data.  If we use the power spectrum of Veeraraghavan \& Davies (1997) for free-free emission and assume that the other component is independent of angular scale, as observations to date seem to roughly indicate, or if we use the 12 $\mu$m template as a tracer of this component, then we find that from spectral index considerations this component could be spinning dust emission.  Other CMB data sets would have to be studied to confirm this.  Given that the two components may have different power spectra, which component dominates in a given data set would depend on both the frequency as well as the relevant angular scales.  Higher quality CMB anisotropy data, more knowledge about the characteristics of dust emission at high frequencies and at microwave frequencies, and high resolution $H_{\alpha}$ data over large portions of the sky, will all be valuable in obtaining a clearer picture of foreground contaminants in low frequency microwave data.

We are indebted to D. Finkbeiner for valuable and detailed comments on the manuscript.  We acknowledge useful discussions with R. Kneissl, J. Macias-Perez, R. Reynolds, and T. Souradeep, and the use of the WCS Tools software (Mink 1997).  We thank D. Finkbeiner for the 12 $\mu$m dust template, J. Gaustad and P. McCullough for the SHASSA $H_{\alpha}$ data (SHASSA was produced with support from the NSF), and D. Mink for assistance with coordinate fitting in the VTSS $H_{\alpha}$ observations, near a singularity in the celestial coordinate system.  We thank the referee for very useful comments that helped improve the manuscript.  BD and JHS acknowledge support from  NSF grants AST-9800476 and AST-0098487 and a Horton Foundation grant to Virginia Tech. The Miles C. Horton, Sr. $\!\!$Research Center is operated by the Virginia Polytechnic Institute and State University with support from the Horton Foundation.  This work was partially carried out at the Infrared Processing and Analysis Center and the Jet Propulsion Laboratory of the California Institute of Technology, under a contract with the National Aeronautics and Space Administration.  PM and BR acknowledge support from NSF CAREER grant AST-9875031.

\clearpage
\begin{small}
\begin{deluxetable}{lccc}
\tablewidth{0pt}
\tablecaption{Dust Temperatures and 100 $\mu$m Template Intensities}
\tablehead{
\colhead{Region} & 
\colhead{Average Dust Temperature (K)} & 
\colhead{i100$^{\tablenotemark{a}}$ (MJy/sr)} & 
\colhead{cdi100$^{\tablenotemark{b}}$ (MJy/sr)}
}
\startdata
OVRO Fields & 16.2 & 5.90 & 16.1 \\
NCP (DEC $>81\dg$) & 16.9 & 4.08 & 8.41 \\
$|b|>30\dg$ & 17.9 & 2.12 & 2.56 \\
\enddata
\tablenotetext{a}{\small{The 100 $\mu$m dust emission template.}}
\tablenotetext{b}{\small{The 100 $\mu$m column density template that assumes a constant dust temperature of 18 K.}}
\end{deluxetable}

\begin{deluxetable}{lccccccc}
\tablecolumns{11} 
\tablewidth{0pt}
\tablecaption{Results of Correlating the OVRO Data with Foreground Templates$^{\tablenotemark{a}}$}
\tablehead{
\colhead{} & 
\multicolumn{3}{c}{14.5 GHz} & \colhead{} &
\multicolumn{3}{c}{31.7 GHz} \\
\cline{2-4} \cline{6-8} \\ [-3mm] 
\colhead{Template} &
\colhead{$a\pm \delta a^{\tablenotemark{b}}$} &
\colhead{$\frac{a}{\delta a}$} & \colhead{$\Delta T$ $(\mu K)$} &\colhead{} &
\colhead{$a\pm \delta a^{\tablenotemark{b}}$} &
\colhead{$\frac{a}{\delta a}$} & \colhead{$\Delta T$ $(\mu K)$} 
}
\startdata 
cdi100$^{\tablenotemark{c}}$ & $50.8\pm 5.3$ & 9.6 & $134\pm14.0$ && $6.6\pm 5.3$ & 1.2 & $17.3\pm 14.0$ \\
i100$^{\tablenotemark{d}}$ & $145\pm 15.7$ & 9.2 & $136\pm14.8$ &&$19.5\pm 15.2$ & 1.3 & $18.3\pm 14.3$ \\
i60$^{\tablenotemark{e}}$ & $1210\pm 133$ & 9.1 & $138\pm15.1$ &&$191\pm 128$ & 1.5 & $21.7\pm 14.6$ \\
i12$^{\tablenotemark{f}}$ & $2190\pm 373$ & 5.9 & $91.6\pm15.6$ && $578\pm 361$ & 1.6 & $24.1\pm 15.1$\\ \hline
i12 & $-246\pm 506$ & $-0.5$ & $-10.0\pm 20.6$ && $484\pm 480$ & 1.0 & $20.2\pm 20.1$ \\
i100 & $152\pm 21.3$ & 7.1 & $143\pm 20.1$ && $6.0\pm 20.3$ & 0.3 & $5.6\pm 19.1$ \\[2mm]
i60 & $531\pm 336$ & 1.6 & $60.5\pm 38.2$ && $245\pm 312$ & 0.8 & $27.9\pm 35.5$ \\
i100 & $87.3\pm 39.6$ & 2.2 & $82.1\pm 37.3$ && $-7.1\pm 37.1$ & $-0.2$ & $-6.7\pm 34.9$ \\ \hline
$H_{\alpha}$ & $-17.0\pm19.6$ & $-0.9$ & $-15.7\pm18.0$ && $-2.0\pm17.9$ & $-0.1$ & $-1.7\pm14.8$ \\[2.mm]
$H_{\alpha}$ & $-36.0\pm19.7$ & $-1.8$ & $-29.9\pm16.4$ && $-5.3\pm18.0$ & $-0.3$ & $-4.4\pm15.0$ \\
i100 \negthinspace& $148\pm15.8$ & 9.4 & $139\pm14.9$ && $20.1\pm15.4$ & 1.3 & $18.9\pm14.5$\\[2.mm]
$H_{\alpha}$ \negthinspace& $-70.5\pm21.1$ & $-3.3$ & $-58.7\pm17.6$ && $-18.6\pm20.0$ & $-0.9$ & $-15.5\pm16.6$ \\
i12 & $2700\pm403$ & 6.7 & $110\pm16.3$ && $747\pm404$ & 1.8 & $30.3\pm16.4$ \\                 
\enddata
\vspace{-0.2in}
\tablenotetext{a}{\small{The upper panel in this table refers to individual correlations of the OVRO data with dust templates, the central panel considers joint correlations of the data with dust templates and the bottom panel refers to correlations of th
e data including the $H_{\alpha}$ template.}}
\tablenotetext{b}{\small{Correlation slope $a$ in units of $\mu$K(MJy/sr)$^{-1}$ for microwave-dust correlations, and in units of $\mu$K(R)$^{-1}$ for microwave-$H_{\alpha}$ correlations.}}
\tablenotetext{c}{\small{The 100 $\mu$m column density template that assumes a constant dust temperature of 18 K.}}
\tablenotetext{d}{\small{The 100 $\mu$m dust emission template.}}
\tablenotetext{e}{\small{The 60 $\mu$m dust emission template.}}
\tablenotetext{f}{\small{The 12 $\mu$m dust emission template.}}
\end{deluxetable}
\end{small}

\begin{landscape}
\begin{deluxetable}{lcccccccccccccc} 
\tablecolumns{15} 
\tablewidth{0pc} 
\tablecaption{Results from Combining Channels of the SP94 Data\tablenotemark{a}}
\tablehead{
\colhead{}    &  \multicolumn{4}{c}{$K_a$} & \colhead{} &  \multicolumn{4}{c}{$Q$} & \colhead{} &  \multicolumn{4}{c}{$K_a+Q$} \\ 
\cline{2-5} \cline{7-10} \cline{12-15}\\
\colhead{Template} & \colhead{$n$} & \colhead{$a\pm \delta a^{\tablenotemark{b}}$} & \colhead{$\frac{a}{\delta a}$} & \colhead{$\Delta T^{\tablenotemark{c}}$} &\colhead{} & \colhead{$n$} & \colhead{$a\pm \delta a^{\tablenotemark{b}}$} &  \colhead{$\frac{a
}{\delta a}$} & \colhead{$\Delta T^{\tablenotemark{c}}$} &\colhead{} & \colhead{$n$} & \colhead{$a\pm \delta a^{\tablenotemark{b}}$} &  \colhead{$\frac{a}{\delta a}$} & \colhead{$\Delta T^{\tablenotemark{c}}$} \\
}
\startdata 
i100 & 0.0 & $28.9\pm 32.3$ & 0.9 & $8.5\pm 11.3$ && 0.0 & $62.6\pm 42.7$ & 1.5 & $23.4\pm 16.1$ && 0.0 & $24.8\pm 34.4$ & 0.7 & $5.0\pm 12.1$\\
  & ${\tablenotemark{d}}$... & ... & ... & ... && 0.8 & $63.1\pm 42.3$ & 1.5 & $23.9\pm 16.0$ && 2.2 & $53.4\pm 22.8$ & 2.3 & $20.8\pm 8.9$ \\[2.mm]
i12 & 0.0 & $129\pm 716$ & 0.2 & $-0.2\pm 11.5$ && 0.0 & $1150\pm 799$ & 1.4 & $20.1\pm 14.4$ && 0.0 & $431\pm 756$ & 0.6 & $0.7\pm 12.2$ \\
  & ${\tablenotemark{d}}$... & ... & ... & ... && 3.2 & $1190\pm 697$ & 1.7 & $22.1\pm 12.9$ && 3.0 & $854\pm 359$ & 2.4 & $16.2\pm 6.8$ \\
\enddata 
\tablenotetext{a}{\small{We combine the $K_a$ band channels, the $Q$ band channels, and finally all the channels, using first a spectral index $n=0$ and then the best fit $n$ for the case when the SP94 data are correlated with the 100 $\mu$m and 12 $\mu$m
 templates individually.  Here $K_a$ band results use a reference frequency $\nu_0=30$ GHz, $Q$ band results use $\nu_0=42$ GHz, and when all channels are combined together results are shown for $\nu_0=36$ GHz.  Only 39 of 43 pixels for each channel were 
considered in this analysis.}}
\tablenotetext{b}{\small{Correlation slope $a$ in units of $\mu$K(MJy/sr)$^{-1}$.}}
\tablenotetext{c}{\small{Temperature anisotropy $\Delta T$ in units of $\mu$K.}}
\tablenotetext{d}{\small{In this case the best fit $n$ was greater than 15, outside the range of the $n$ grid used.}}
\end{deluxetable}
\end{landscape}

\clearpage
\begin{figure}
\small{
\centerline{\epsfig{
file=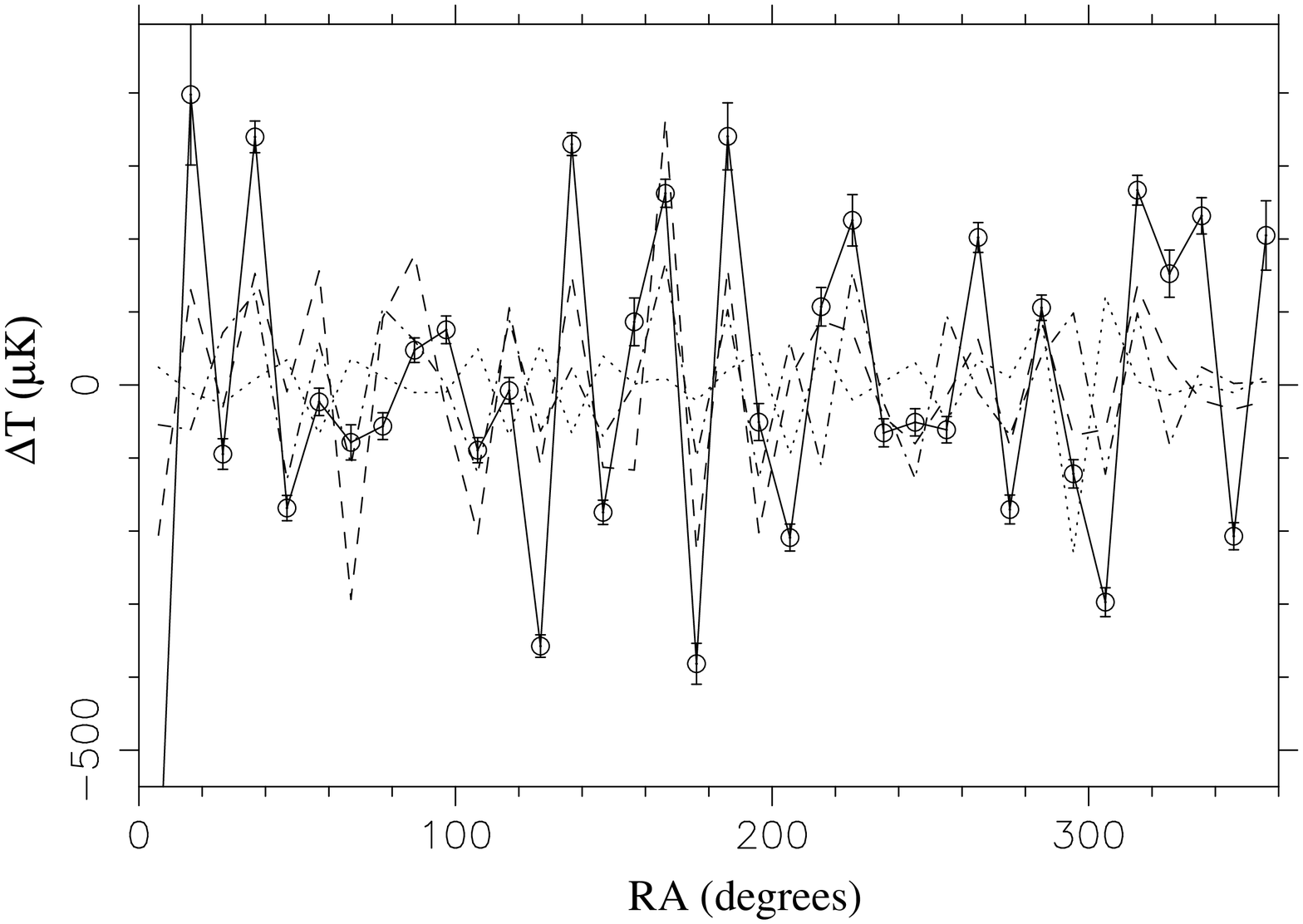,width=14cm,angle=0}}
\caption{The figure shows the OVRO data (open circles connected by solid lines), the 100 $\mu$m-correlated emission (dashed line), the 12 $\mu$m-correlated emission (dot-dashed line) and the $H_{\alpha}$-correlated emission (dotted line), all at 14.5 GHz.
}
\label{fig1}}
\end{figure}

\begin{figure}
\small{
\centerline{\epsfig{
file=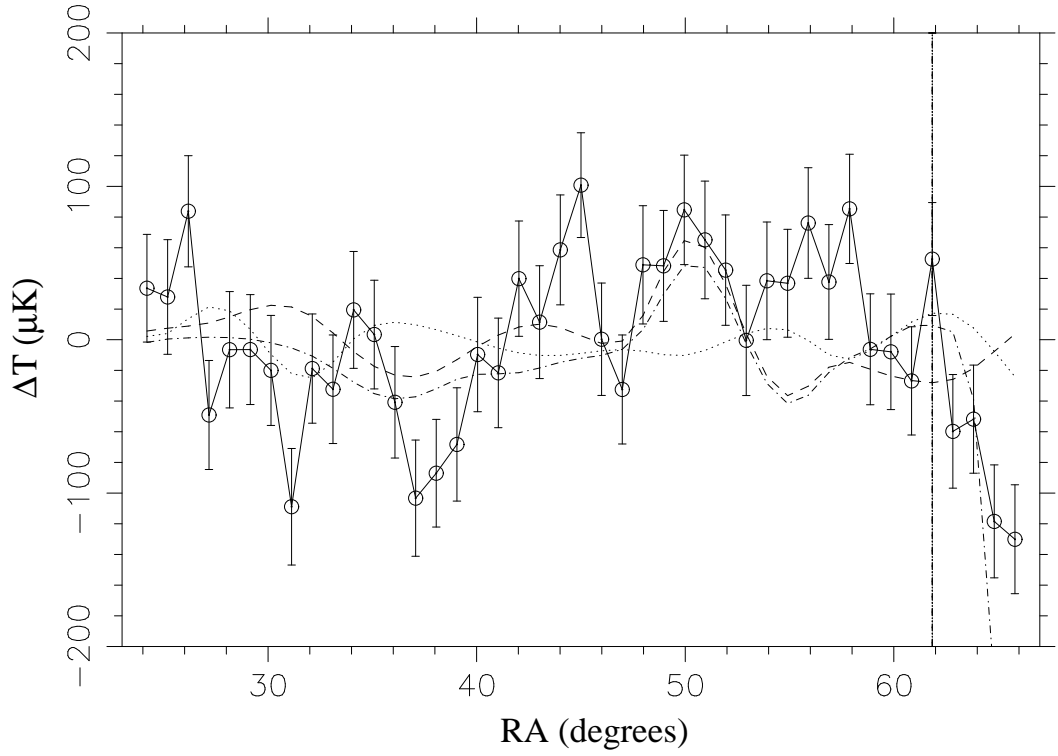,width=14cm,angle=0}}
\caption{The figure shows the SP94 $Q$ band coadded data (open circles connected by solid lines), the 100 $\mu$m-correlated emission (dashed line), the 12 $\mu$m-correlated emission (dot-dashed line) and the $H_{\alpha}$-correlated emission (dotted line) in the $Q$ band. The vertical line is at the 39th data point marking out the 4 data points that have not been used in this analysis (see text).}
\label{fig2}}
\end{figure}

\begin{figure}
\small{
\centerline{\epsfig{
file=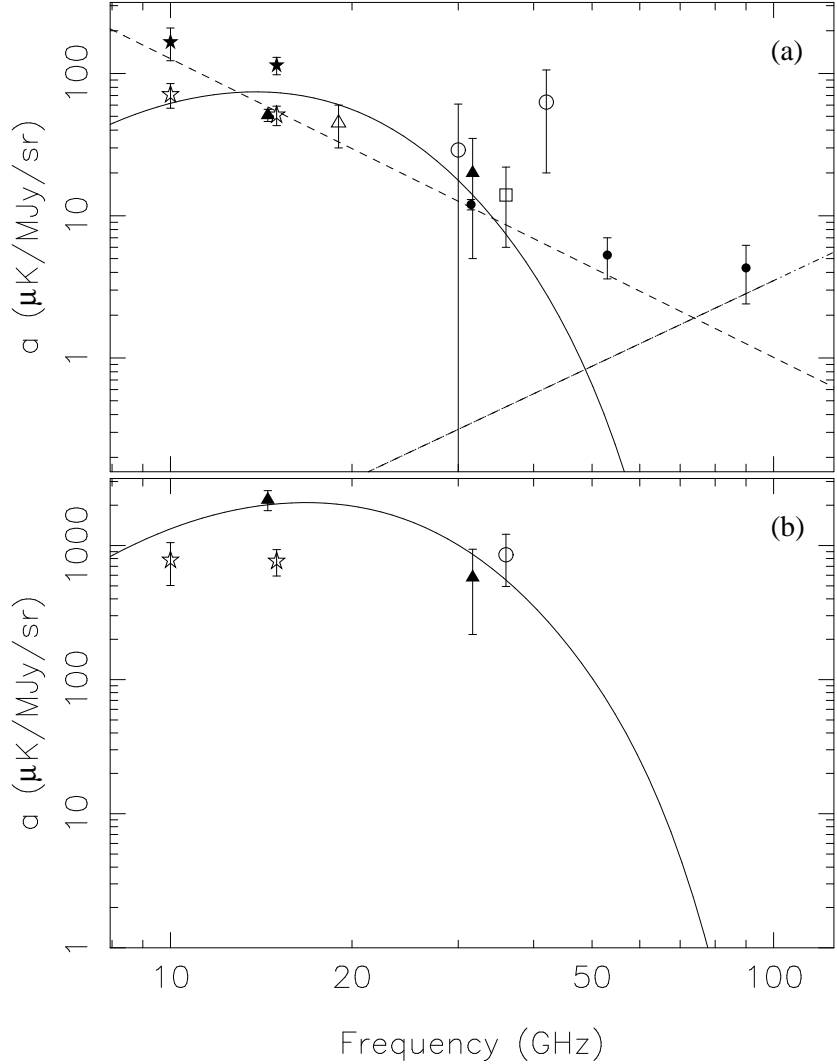,width=11cm,angle=0}}
\caption{(a) 100 $\mu$m-correlated emission in the Tenerife (open stars, de Oliveira-Costa et al. 2001; filled stars, Mukherjee et al. 2001), OVRO (filled triangles), 19 GHz (open triangles), SP94 (open circles), Saskatoon (open square) and DMR (filled circles) microwave data. (b) 12 $\mu$m-correlated emission reported so far in microwave data (symbols are the same as above).  The solid curve is representative of a spinning dust spectrum, the dashed and the dot-dashed lines represent free-free and vibrational dust emission spectrums respectively.}
\label{fig3}}
\end{figure}


\clearpage
\begin{thebibliography}{99}
\bibitem[Banday et al. (1996)]{banday}
Banday, A.J., G\'orski, K.M., Bennett, C.L., Hinshaw, G., Kogut, A., \& Smoot, G.F. 1996, ApJ, 468, L85 
\bibitem[Barreiro, \& Hobson (2001)]{barreiro}Barreiro, R.B., \& Hobson, M. 2001, MNRAS, 327, 813
\bibitem[Bartlett, \& Amram (1998)]{bartlett}Bartlett, J.G., \& Amram, P. 1998, astro-ph/9804330
\bibitem[Chamberlain 1961]{Ch} Chamberlain, J.W. 1961, Physics of the Airglow and Aurora (New York: Academic)
\bibitem[Coble et al. (1999)]{coble}Coble, K., et al. 1999, ApJ, 519, L5 
\bibitem[Chiang, \& Coles (2000)]{chiang}Chiang, L.Y., \& Coles, P. 2000, MNRAS, 311, 809
\bibitem[Dennison, Topasna, \& Simonetti (1997)]{dts} Dennison, B., Topasna, G.A., \& Simonetti, J.H. 1997, \apjl, 474, L31
\bibitem[de Oliveira-Costa et al. (1997)]{costa97}de Oliveira-Costa, A., et al. 1997, ApJ, 482, L17
\bibitem[de Oliveira-Costa et al. (2000)]{costa00}de Oliveira-Costa, A., et al. 2000, ApJ, 542, L5
\bibitem[de Oliveira-Costa et al. (2002)]{costa02}de Oliveira-Costa, A., et al. 2002, ApJ, 567, 363
\bibitem[de Oliveira-Costa et al. (1999)]{costa99}de Oliveira-Costa, A., Tegmark, M., Gutierrez, C.M., Jones, A.W., Davies, R.D., Lasenby, A.N., Rebolo, R. \& Watson, R.A. 1999, ApJ, 527, L9
\bibitem[de Oliveira-Costa et al. (1998)]{costa98}de Oliveira-Costa, A., Tegmark, M., Page, L.A, \& Boughn, S.P. 1998, ApJ, 509, L9
\bibitem[Draine, \& Lazarian (1998a)]{draine98a}
Draine, B.T., \& Lazarian, A. 1998a, ApJ, 494, L19
\bibitem[Draine, \& Lazarian (1998b)]{draine98b}
Draine, B.T., \& Lazarian, A. 1998b, ApJ, 508, 157
\bibitem[Draine, \& Lazarian (1999)]{draine99}
Draine, B.T., \& Lazarian, A. 1999, in ASP Conf. Ser. 181, Microwave Foregrounds, ed. A. de Oliveira-Costa \& M. Tegmark (San Francisco: ASP), 133
\bibitem[Efstathiou]{Efstathiou01}Efstathiou, G., et al. 2001, MNRAS, 330L, 29
\bibitem[Finkbeiner, \& Schlegel (1999)]{fink99}Finkbeiner, D.P., \& Schlegel, D.J. 1999, in ASP Conf. Ser. 181, Microwave Foregrounds, ed. A. de Oliveira-Costa \& M. Tegmark (San Francisco: ASP), 101
\bibitem[Finkbeiner, Schlegel, \& Frank (2002)]{fink02}Finkbeiner, D.P., Schlegel, D.J., Frank, C. \& Heiles, C. 2002, ApJ, 566, 898
\bibitem[Ganga et al. (1997)]{ganga97}Ganga, K., Ratra, B., Gundersen, J.O., \& Sugiyama, N. 1997, ApJ, 484, 7
\bibitem[Ganga et al. (1998)]{ganga98}Ganga, K., Ratra, B., Lim, M.A., Sugiyama, N., \& Tanaka, S.T. 1998, ApJS, 114, 165
\bibitem[Gaustad et al. (2001)]{gaustad01}Gaustad, J.E., McCullough, P.R., Rosing, W., \& Van Buren, D. 2001, PASP, 113, 1326
\bibitem[Gaustad, McCullough, \& Van Buren (1996)]{gaustad96}Gaustad, J.E., McCullough, P.R., \& Van Buren, D. 1996, PASP, 108, 351
\bibitem[Gorski et al. (1996)]{gorski96}
G\'orski, K.M., Banday, A.J., Bennett, C.L., Hinshaw, G., Kogut, A., Smoot, G.F., \& Wright, E.L. 1996, ApJ, 464, L11
\bibitem[Gundersen et al. (1995)]{gundersen95}Gundersen, J.O., et al. 1995, ApJ, 443, L57
\bibitem[Hamilton, \& Ganga (2001)]{hamilton}Hamilton, J.-Ch., \& Ganga K.M. 2001, A\&A, 368, 760
\bibitem[Knox, \& Page (2000)]{knox00} Knox, L., \& Page, L. 2000, Phys. Rev.
   Lett., 85, 1366 
\bibitem[Kogut (1997)]{kogut97}Kogut, A. 1997, AJ, 114, 3
\bibitem[Kogut (1999)]{kogut99}Kogut, A. 1999, in ASP Conf. Ser. 181, Microwave Foregrounds, ed. A. de Oliveira-Costa \& M. Tegmark (San Francisco: ASP), 91
\bibitem[Kogut et al. (1996a)]{kogut96a}Kogut, A., Banday, A.J., Bennet, C.L., G\'orski, K.M., Hinshaw, G. \& Reach, W.T. 1996a, ApJ, 460, 1
\bibitem[Kogut et al. (1996b)]{kogut96b}Kogut, A., Banday, A.J., Bennet, C.L., G\'orski, K.M., Hinshaw, G., Smoot, G.F. \& Wright, E.L. 1996b, ApJ, 464, L5
\bibitem[Lagache et al. (2000)]{lagache}Lagache, G., Haffner, L.M., Reynolds, R.J., \& Tufte S.L. 2000, A\&A, 354, 247
\bibitem[Leitch et al. (1997)]{leitch97}Leitch, E.M., Readhead, A.C.S., Pearson, T.J., \& Myers, S.T. 1997, ApJ, 486, L23
\bibitem[Leitch et al. (2000)]{leitch00}Leitch, E.M., Readhead, A.C.S., Pearson, T.J., Myers, S.T., Gulkis, S., \& Lawrence C.R. 2000, ApJ, 532, 37
\bibitem[Lim et al. (1996)]{lim}Lim, M.A., et al. 1996, ApJ, 469, L69 
\bibitem[McCullough (1997)]{mc97}McCullough, P.R. 1997, ApJ, 113, 6
\bibitem[McCullough et al. (1999)]{mc99}McCullough, P.R., Gaustad, J.E., Rosing, W., \& Van Buren, D. 1999, astro-ph/9902248
\bibitem[Mink (1997)]{mink}Mink, D.J. 1997, in ASP Conf. Ser. 125, Astronomical Data Analysis Software and Systems VI, ed. G. Hunt and H. E. Payne (San Francisco: ASP), 249
\bibitem[Mukherjee et al. (2001)]{mukherjee}Mukherjee, P., Jones, A.W., Kneissl, R., \& Lasenby, A.N. 2001, MNRAS, 320, 224
\bibitem[Netterfield et al. (2002)]{netterfield02}Netterfield, C.B., et al. 2002, \apj, 571, 604
\bibitem[Park et al. (1998)]{park98}Park, C., Colley, W.N., Gott, J.R., Ratra, B., Spergel, D.N., \& Sugiyama, N. 1998, ApJ, 506, 473
\bibitem[Park et al. (2002)]{park02}Park, C.-G., Park, C., \& Ratra, B. 2002, ApJ, 568, 9
\bibitem[Park et al. (2001)]{park01}Park, C.-G., Park, C., Ratra, B., \& Tegmark, M. 2001, ApJ, 556, 582
\bibitem[Podariu et al. (2001)]{podariu} Podariu, S., Souradeep, T., Gott, J.R., Ratra, B., \& Vogeley, M.S. 2001, ApJ, 559, 9
\bibitem[Pryke et al. (2002)]{pryke}Pryke, C., Halverson, N.W., Leitch, E.M., Kovac, J., Carlstrom, J.E., Holzapfel, W.L., \& Dragovan, M. 2002, \apj, 568, 46
\bibitem[Ratra et al. (1999a)]{ratra99a} Ratra, B., Ganga, K., Stompor, R., Sugiyama, N., de Bernardis, P., \&\ G\'orski, K.M. 1999a, ApJ, 510, 11
\bibitem[Ratra et al. (1999b)]{ratra99b} Ratra, B., Stompor, R., Ganga, K., Rocha, G., Sugiyama, N., \&\ G\'orski, K.M. 1999b, ApJ, 517, 549 
\bibitem[Ratra et al. (1997)]{ratra97} Ratra, B., Sugiyama, N., Banday, A.J., \& G\'orski, K.M. 1997, \apj, 481, 22
\bibitem[Rengelink et al. (1997)]{rengelink}Rengelink, R.B., Tang, Y., Debruyn, A.G., Miley, G.K., Bremer, M.N., Rottgering, H.J.A., \& Bremer, M.A.R. 1997, A\&AS, 124, 259
\bibitem[Reynolds, \& Haffner (2002)]{reynolds02}Reynolds, R.J., \& Haffner, L.M. 2002, PASP, in press 
\bibitem[Reynolds et al. (1995)]{reynolds95}Reynolds, R.J., Tufte, S.L., Kung, D.T., McCullough, P.R., \& Heiles, C. 1995, ApJ, 448, 715
\bibitem[Rocha et al. (1999)]{rocha99a} Rocha, G., Stompor, R., Ganga, K.,  Ratra, B., Platt, S.R., Sugiyama, N., \& G\'orski, K.M. 1999, ApJ, 525, 1
\bibitem[Schlegel, Finkbeiner, \& Davis (1998)]{schlegel98}Schlegel, D. J., Finkbeiner, D.P., \& Davis, M. 1998, ApJ, 500, 525
\bibitem[Shandarin et al. (2002)]{shandarin}Shandarin, S.F., Feldman, H.A., Xu, Y., \& Tegmark, M. 2002, ApJS, 141, 1
\bibitem[Simonetti, Dennison, \& Topasna (1996)]{sdt}Simonetti, J.H., Dennison, B., \& Topasna G.A. 1996, ApJ, 458, L1 
\bibitem[Stompor et al. (2001)]{stompor}Stompor, R., et al. 2001, \apj, 561, L7
\bibitem[Valls-Gabaud (1998)]{vg}Valls-Gabaud, D. 1998, PASP, 15, 111
\bibitem[Veeraraghavan, \& Davies (1997)]{vd}Veeraraghavan, S., \& Davies, R.D. 1997, unpublished manuscript
\bibitem[Wang, \& Mathews (2002)]{wang00}Wang, Y., \& Mathews, G. 2002, \apj, 573, 1
\bibitem[Wang (2001)]{wang01}Wang, X., Tegmark, M., \& Zaldarriaga, M. 2002, Phys. Rev. D 65, 123001
\bibitem[Weingartner, \& Draine (2000)]{wd}Weingartner, J.C., \& Draine, B.T. 2000, ApJ, 553, 581
\bibitem[Wu et al. (2001)]{wu}Wu, J.-H.P., et al. 2001, Phys. Rev. Lett., 87, 251303
\end{thebibliography}
\end{document}